\documentclass[12pt,epsf]{article}
\usepackage{amssymb,amsmath,amsbsy}
\usepackage{graphicx}
\usepackage{setspace}
\usepackage{subfigure}
\usepackage{cite}
\usepackage{customcommands}

\makeatletter
\def\namedlabel#1#2{\begingroup
    #2%
    \def\@currentlabel{#2}%
    \phantomsection\label{#1}\endgroup
}
\makeatother

\usepackage{color}

\usepackage{xcolor}
\usepackage[citebordercolor=green, linkbordercolor={ 0 0 1}, linktocpage=true]{hyperref}

\begin{document}
\begin{titlepage}
\begin{NoHyper}
\hfill
\vbox{
    \halign{#\hfil         \cr
           } 
      }  
\vspace*{20mm}
\begin{center}
{\Large \bf Locality from Quantum Gravity: All or Nothing}

\vspace*{1.6cm}
Netta Engelhardt$^a$ and Sebastian Fischetti$^b$\\

\vspace*{1cm}
\let\thefootnote\relax\footnote{nengelhardt@princeton.edu, s.fischetti@imperial.ac.uk}

{$^{a}$Department of Physics, Princeton University\\
Princeton, NJ 08544, USA\\
\vspace{0.25cm}
$^{b}$Theoretical Physics Group, Blackett Laboratory, Imperial College \\ London SW7 2AZ, UK}

\vspace*{2cm}
\end{center}
\begin{abstract}
In a full theory of quantum gravity, local physics is expected to be approximate rather than innate. It is therefore important to understand how approximate locality emerges in the semiclassical limit.  Here we show that any notion of locality emergent from a holographic theory of quantum gravity is ``all or nothing'': local data is not obtained gradually from subregions of the boundary, but is rather obtained all at once when enough of the boundary is accessed. Our assumptions are mild and thus this feature is quite general; in the special case of AdS/CFT, a slightly different manifestation follows from well-known and familiar properties.

\vfill
{\noindent \footnotesize Written for the Gravity Research Foundation 2017 Awards for Essays on Gravitation}
\end{abstract}
\end{NoHyper}

\end{titlepage}
\vskip 1cm
\begin{spacing}{1.2}

\setcounter{footnote}{0}


Recent developments suggest that local -- that is, semiclassical -- gravitational physics is not fundamental, but rather emerges from an appropriate limit of a complete theory of quantum gravity.  A mysterious aspect of this emergence is its rarity: generic states of quantum gravity are non-classical. As our own universe is one of the special semiclassical states, it is of vital importance to understand how a theory of quantum gravity gives rise to local physics. 

What precisely do we mean by local physics? An ordinary quantum field theory is local if the commutator of any field $\phi$ at two spacelike-separated points $x$ and $y$ vanishes: $[\phi(x),\phi(y)]=0$. This concept is only well-defined in a system that has at least \textit{(i)} an approximate notion of points and \textit{(ii)} the information to determine the causal separation between the two points (whether they are spacelike-, timelike-, or null-separated). These two data constitute a semiclassical \textit{conformal geometry}, a more primitive and coarser construct than a full semiclassical geometry: the latter contains the data required to measure distances between points, which is absent in the former. Understanding the emergence of locality is thus tantamount to understanding the emergence of a semiclassical conformal geometry.

Because we are ultimately interested in describing our own universe, we restrict our attention to semiclassical (conformal) spacetimes containing matter.  Specifically, we make the following assumption:
\begin{description}
	\item[\namedlabel{A1}{A1}] A semiclassical spacetime contains a weakly interacting quantum matter field $\phi(x)$.
\end{description}
While this assumption is a reasonable expectation of any putative theory of quantum gravity, little else is known explicitly about such theories.  Fortunately, the holographic principle~\cite{Tho93, Sus95, Bou02} provides valuable insights: any theory of quantum gravity in~$(d+1)$ dimensions is believed to be expressible \textit{indirectly} in terms of a non-gravitational theory in~$d$ dimensions; typically these theories are termed the ``bulk'' and ``boundary'', respectively.  In this Essay, we draw from our earlier work in~\cite{EngFis17} to show that the holographic principle implies that the emergence of semiclassical locality from a holographic theory of quantum gravity obeys a key property: it is ``all or nothing''.

To facilitate our arguments, we first make our assumptions about holography explicit:
\begin{description}
	\item[\namedlabel{A2}{A2}] The boundary theory lives on a geometry that can be embedded as a timelike or null hypersurface in the bulk;
	\item[\namedlabel{A3}{A3}] For each~$n$, the boundary contains an object~$O_n(X_1,\ldots,X_n)$ which in the limit of a semiclassical bulk is related to the~$n$-point correlator of~$\phi(x)$ as
	\be \label{eq:dict}
	\lim_{x_i \to X_i} \ev{\phi(x_1) \cdots \phi(x_n)} = O_n(X_1,\ldots,X_n),
	\ee
	where~$x_i$ and~$X_i$ label points in the bulk and boundary, respectively.
\end{description}
While the reader may protest that \ref{A3} seems draconianly restrictive, it is in fact rather mild: for timelike separated $X_{i}$, the left hand side of~\eqref{eq:dict} is morally the S-matrix of the bulk theory. Any boundary theory should contain an object that encodes bulk scattering data if it is to describe the bulk.

\begin{figure}[t]
\centering
\includegraphics[width=0.3\textwidth,page=1]{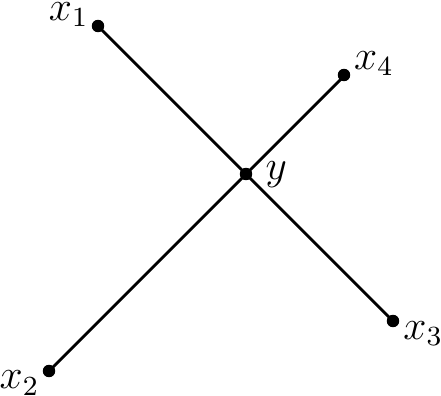}
\caption{When the points~$x_i$ are null-separated from a vertex~$y$ at which energy-momentum is conserved, the correlator~$\ev{\phi(x_1) \cdots \phi(x_n)}$ is singular.}
\label{fig:Landau}
\end{figure}

\begin{figure}[t]
\centering
\includegraphics[page=2,width=0.25\textwidth]{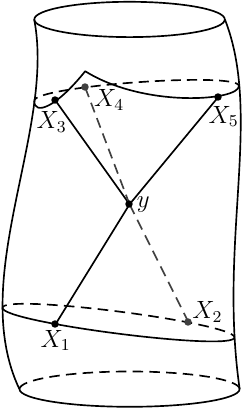}
\caption{The singularities of the boundary object~$O_n(X_i)$ can be used to pick out a bulk point~$y$ when all the~$X_i$ are null-separated from~$y$; this allows the construction of the lightcone cuts of~$y$ as the intersection of the lightcone of~$y$ with the boundary.}
\label{fig:cut}
\end{figure}

The three ingredients~\ref{A1}-\ref{A3} can now be combined.  First, an immediate consequence of~\ref{A1}: if the bulk is semiclassical, the correlator~$\ev{\phi(x_1) \cdots \phi(x_n)}$ (taking~$n \geq 4$) is singular when the~$x_i$ are null-separated from a common vertex~$y$, as long as the corresponding position-space Feynman diagram (shown in Figure~\ref{fig:Landau}) conserves energy-momentum at~$y$~\cite{MalSim15} (see also earlier work by~\cite{PolSus99, GarGid09, HeePen09, Pen10, OkuPen11}). These so-called lightcone singularities -- in which the~$x_i$ are distinct and some are timelike separated -- are therefore sensitive to the causal structure of the bulk.  It then follows from~\ref{A2} and \ref{A3} that whenever the bulk is well-approximated by a semiclassical conformal geometry, the \textit{boundary} object~$O_n(X_i)$ is singular when the boundary points~$X_i$ are null-separated from a \textit{bulk} point~$y$ (at which energy-momentum is conserved)\footnote{In the context of AdS/CFT, these were termed ``bulk-point singularities'' in~\cite{MalSim15}.}; see Figure~\ref{fig:cut}.   Thus we immediately obtain a necessary condition for the emergence of a semiclassical conformal geometry: the~$O_n(X_i)$ must exhibit singularities when the~$X_i$ are distinct and at least some are timelike separated. 

Bulk locality plays an essential role in the above observation, as these singularities in~$O_n(X_i)$ may be traced to the shared vertex~$y$, which uniquely identifies a bulk point.  It is therefore instructive to ask whether this argument can be reversed: that is, is a semiclassical bulk causal structure encoded in the singularities of~$O_n(X_i)$?  As shown in~\cite{EngHor16a,EngHor16c}, the answer is yes.  Because the singularities of~$O_n(X_i)$ correspond to null-separation from a bulk point~$y$, they can be used to construct spacelike slices of the boundary corresponding to the intersection of lightcones of bulk points with the boundary; these so-called ``lightcone cuts'' are shown in Figure~\ref{fig:cut}.  Two such cuts are tangent if and only if the corresponding bulk points are null-separated; this feature implies that \textit{the geometric structure of the cuts represents the bulk conformal geometry}.

The upshot is that if the singularity structure of~$O_n(X_i)$ defines cuts that give rise to a consistent conformal geometry via the procedure of~\cite{EngHor16a,EngHor16c}, then that \textit{is} the emergent bulk dual.  Manifest in this construction is the emergence of locality: the very notion of a bulk point~$y$ is contained in the singularity structure of~$O_n(X_i)$ for some boundary points~$X^{(y)}_i$ null-separated from~$y$, and the conformal geometry in a neighborhood of~$y$ is contained in the singularity structure of~$O_n(X_i)$ in a neighborhood of~$X^{(y)}_i$.  We have thus arrived at a sufficient and necessary condition for the emergence of locality from a state of holographic quantum gravity: the object~$O_n(X_i)$ must feature singularities on boundary spatial slices, and the geometry of those spatial slices must give rise via the procedure of~\cite{EngHor16a,EngHor16c} to a consistent semiclassical conformal geometry.

\newpage

\subsubsection*{All or Nothing}

Can we use this precise correspondence to understand properties of emergent bulk locality?

\begin{figure}[t]
\centering
\subfigure[]{\includegraphics[page=3,width=0.23\textwidth]{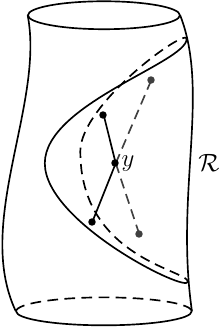}
\label{subfig:big}
}
\hspace{2cm}
\subfigure[]{\includegraphics[page=4,width=0.23\textwidth]{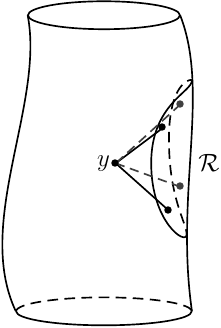}
\label{subfig:small}
}
\caption{In order to obtain a bulk point~$y$ from singularities of~$O_n(X_i)$, the boundary points~$X_i$ need to be null-separated from~$y$ and sufficiently spread out to conserve momentum at~$y$.  If~$O_n(X_i)$ is restricted to some subregion~$\Rcal$ of the boundary,~$y$ can only be reconstructed if~$\Rcal$ is sufficiently large, as in~\subref{subfig:big}; if~$\Rcal$ is too small, it cannot, as in~\subref{subfig:small} where there is no ``left-directed'' momentum.}
\label{fig:spreadout}
\end{figure}

The above construction of a bulk point~$y$ (and the conformal geometry at~$y$) from singularities of the object~$O_n(X_i)$ relies on lightcone singularities; these exist only if \textit{(i)} the $X_i$ are all null-separated from a common vertex $y$, and \textit{(ii)} energy-momentum is conserved at $y$.  This latter condition implies that the construction of $y$ requires access to a ``sufficiently spread-out'' set of boundary points~$X_i$, as shown in Figure~\ref{fig:spreadout}.  A dramatic consequence is immediate: if the $X_i$ are restricted to lie in too small a region $\Rcal$, then~$O_n(X_i)$ may not be singular even if the $X_i$ are null separated from $y$; thus it is impossible to identify $y$ from data in $\Rcal$.  As~$\Rcal$ is enlarged, however, the point~$y$ and the conformal geometry at $y$ can be fully constructed as soon as~$\Rcal$ reaches the critical size necessary for points in $\Rcal$ to conserve momentum at $y$.

The essential feature here is that the point~$y$ is not obtained gradually and little-by-little as~$\Rcal$ is enlarged; rather, it is obtained completely and suddenly as soon as~$\Rcal$ becomes sufficiently large.  This feature follows very generally in any holographic framework obeying properties~\ref{A1}-\ref{A3}; this demonstrates our advertised claim that the reconstruction of (approximate) locality in holographic quantum gravity is ``all-or-nothing''. A holographic description of the theory of quantum gravity in the bulk cannot partially describe local data: it must describe it fully or not at all.

\subsubsection*{An Illustrative Example: AdS/CFT}

\begin{figure}[t]
\centering
\subfigure[]{\includegraphics[page=5]{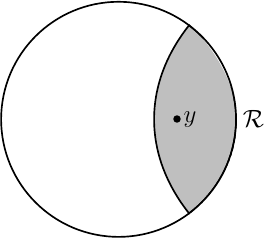}
\label{subfig:bigent}
}
\hspace{2cm}
\subfigure[]{\includegraphics[page=6]{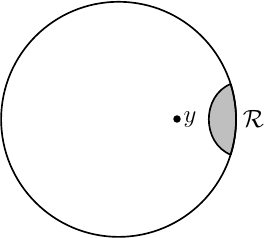}
\label{subfig:smallent}
}
\caption{\subref{subfig:bigent}: In AdS/CFT, bulk operators at a point~$y$ can be obtained from boundary data in a spatial region~$\Rcal$ as long as~$y$ is contained in the RT surface of~$\Rcal$.  \subref{subfig:smallent}: If~$\Rcal$ is too small,~$y$ lies outside the RT surface, and fields at~$y$ cannot be reconstructed from data in~$\Rcal$.}
\label{fig:entanglement}
\end{figure}

This ``all-or-nothing'' emergence is a general feature of holographic frameworks of quantum gravity.  We now consider a concrete example of such a framework: the anti-de Sitter/conformal field theory (AdS/CFT) correspondence~\cite{Mal97, GubKle98, Wit98a}.  This correspondence, which provides the most explicit formulation of holography, relates string (or M) theory with asymptotically AdS boundary conditions to a CFT on the asymptotic boundary of AdS.

Since the asymptotic boundary is timelike, property~\ref{A2} holds.  Moreover, there exists a limit in which the bulk is well-described by semiclassical gravity; in this limit, bulk matter fields are weakly coupled, so property~\ref{A1} holds as well.  Finally, the AdS/CFT dictionary implies that each CFT operator~$\Ocal(X)$ has a dual bulk quantum field~$\phi(x)$ and vice versa, and that the correlators~$\ev{\Ocal(X_1) \cdots \Ocal(X_n)}$ are obtained from the correlators~$\ev{\phi(x_1) \cdots \phi(x_n)}$ as the~$x_i$ are taken to the asymptotic boundary.  Thus property~\ref{A3} holds, with~$O_n(X_1,\ldots,X_n) = \ev{\Ocal(X_1) \cdots \Ocal(X_n)}$.  AdS/CFT is therefore a particular manifestation of the more general holographic framework~\ref{A1}-\ref{A3}.

In fact, access to the full AdS/CFT dictionary, rather than just the limited holographic dictionary of properties~\ref{A2} and~\ref{A3}, shows that the ``all-or-nothing'' property of local bulk data is quite robust.  To see this, recall that a bulk operator~$\phi(y)$ at the point~$y$ can be reconstructed from boundary data in a spatial region~$\Rcal$ if~$y$ lies between the so-called Ryu-Takayanagi (RT) surface of~$\Rcal$ and the boundary~\cite{RyuTak06,DonHar16}, as shown in Figure~\ref{fig:entanglement}.  Thus if~$\Rcal$ is sufficiently small,~$\phi(y)$ cannot be reconstructed from any data in~$\Rcal$; as~$\Rcal$ is enlarged,~$\phi(y)$ (and any other fields at~$y$) can be reconstructed as soon as the RT surface of~$\Rcal$ encloses~$y$.  This feature was interpreted in terms of quantum error correction in~\cite{AlmDon15}, where it was noted that the underlying mechanism is Page's theorem~\cite{Pag93}\footnote{We thank Daniel Harlow for calling our attention to this.}.

Our arguments indicate that the above feature of the AdS/CFT correspondence is a manifestation of the more general principle identified in this Essay: that quantum gravity in general describes locality in an abrupt, global manner, giving access to full data or none, with no continuum in between. In retrospect, this should perhaps have been anticipated: quantization typically induces discretization. There is no reason to expect that a quantum theory of spacetime should produce spacetime in an altogether continuous way; indeed, it does not.

\subsubsection*{Acknowledgements}

We thank Daniel Harlow for comments on an earlier version of this essay.  The work of NE was supported in part by NSF grant PHY-1620059.  SF was supported by STFC grant ST/L00044X/1.

\end{spacing}
\bibliographystyle{jhep}

\bibliography{all}

\end{document}